\documentclass[MSNbibl,nameyear,seceqn,dvips]{arxstspdf}
\usepackage{multirow}
\usepackage{graphicx}
\usepackage{flushend}
\usepackage{stfloats}
\volume{29}
\issue{1}
\pubyear{2014}
\firstpage{18}
\lastpage{25}
\doi{10.1214/13-STS427} 

\begin{document}
\begin{frontmatter}

\title{Bayesian Estimation of Population-Level Trends in Measures of
Health Status} 
\runtitle{Trends in Measure of Health Status}

\begin{aug}
\author[a]{\fnms{Mariel M.} \snm{Finucane}\corref{}\ead[label=e1]{mariel.finucane@gladstone.ucsf.edu}},
\author[b]{\fnms{Christopher J.} \snm{Paciorek}\ead[label=e2]{paciorek@stat.berkeley.edu}},
\author[c]{\fnms{Goodarz} \snm{Danaei}\ead[label=e3]{gdanaei@hsph.harvard.edu}}
\and
\author[d]{\fnms{Majid} \snm{Ezzati}\ead[label=e4]{majid.ezzati@imperial.ac.uk}}
\runauthor{Finucane, Paciorek, Danaei and Ezzati}

\affiliation{University of California, San Francisco, University of
California, Berkeley, Harvard School of Public Health and Imperial
College London}

\address[a]{Mariel M. Finucane is Biostatistician, Gladstone Institutes,
University of San Francisco, California 94158, USA \printead{e1}.}
\address[b]{Christopher J. Paciorek is Associate Research Statistician, Department of Statistics,
University of California, Berkeley 94720, USA \printead{e2}.}
\address[c]{Goodarz Danaei is Assistant Professor, Departments of Epidemiology and of
Global Health and Population,
Harvard School of Public Health, Boston, Massachusetts 02115, USA
\printead{e3}.}
\address[d]{Majid Ezzati is Professor, MRC-HPA Centre for Environment and
Health, Department of Epidemiology and Biostatistics,
School of Public Health, Imperial College London, London W2 1PG, United
Kingdom \printead{e4}.}

\end{aug}

%
\begin{abstract}
Improving health worldwide will require rigorous quantification of
population-level trends in health status. However, global-level
surveys are not available, forcing researchers to rely on fragmentary
country-specific data of varying quality. We present a Bayesian model
that systematically combines disparate data to make country-,
\mbox{region-} and global-level estimates of time trends in important
health indicators.

The model allows for time and age nonlinearity, and it borrows strength
in time, age, covariates, and within and across regional country
clusters to make estimates where data are sparse. The Bayesian approach
allows us to account for uncertainty from the various aspects of
missingness as well as sampling and parameter uncertainty. MCMC
sampling allows for inference in a high-dimensional, constrained
parameter space, while providing posterior draws that allow
straightforward inference on the wide variety of functionals of
interest.

Here we use blood pressure as an example health metric. High blood
pressure is the leading risk factor for cardiovascular disease, the
leading cause of death worldwide. The results highlight a risk
transition, with decreasing blood pressure in high-income regions
and
increasing levels in many lower-income regions.
\end{abstract}

%

\end{frontmatter}

\section{Introduction}
\label{secintro}
Variations and trends in health outcomes and risk factors across the
globe have received greatly increased attention in recent years, in
part driven by the UN's
Millennium Development Goals, the increase in
international funding for global health and the demand for objective
evidence about the effectiveness of interventions. There has been a
concomitant focus on data sources and quantitative methods for
population-level measures of health status. However, global-level
surveys are not available, forcing researchers to rely on fragmentary
country-specific data of varying quality.

The Global Burden of Diseases, Injuries and Risk Factors Study (GBD,
\href{http://www.globalburden.org}{www.globalburden.org}), which aims to quantify the relative
contributions of different diseases and injuries, and their risk
factors, to morbidity and mortality worldwide, offers a demonstration
of these challenges. For example, despite cardiovascular diseases being
the leading causes of death worldwide (\citeauthor{lozano2013},
\citeyear{lozano2013}), our
understanding of their trends is almost entirely based on specific
cohorts and communities, primarily in high-income countries. As part of
the GBD Study, we set out to estimate trends in cardiometabolic risk
factors over the past 30 years for all nations.

In this paper we present a Bayesian model developed to address these
issues by combining disparate data sources to complete the largest-ever
analysis of metabolic risk factors and the first global analysis of
trends. Our model has been used to analyze global trends in systolic
blood pressure (\citeauthor{SBP2010}, \citeyear{SBP2010}), serum
total cholesterol (\citeauthor{TC2010}, \citeyear{TC2010}), body
mass index (\citeauthor{BMI2010}, \citeyear{BMI2010}) and fasting
plasma glucose
(\citeauthor{FPG2011}, \citeyear{FPG2011}).

Here, we focus on the blood pressure analysis as an illustrative example
of model development and the advantages of using the Bayesian
paradigm. 
Kearney et~al. (\citeyear{Kearney2004,Kearney2005}) and \citet{Lawes2004} were
influential in demonstrating the importance of this risk factor, which
is responsible for more than 9 million annual deaths, more than any
other risk factor (\citeauthor{lim2013}, \citeyear{lim2013}). These
analyses, however, were based
on only a small subset of available data. Further, they did not assess
trends over time systematically, did not distinguish
nationally-representative surveys from sub-national and community-based
studies, and did not take into account the missingness of data from
entire countries or age groups when quantifying uncertainty.

In addition to addressing these deficiencies, our approach differs in
important ways from other recent modeling of global health. \citet
{Rajaratnam2010} and \citet{Hogan2010}, for example, modeled global
adult and maternal mortality, respectively. These studies used
investigator-chosen smoothing parameters and implemented a two-stage
estimation procedure, which prevents uncertainty from propagating
through the modeling process. We, on the other hand, estimate all
parameters as part of a single model, allowing all sources of
uncertainty to be reflected in our inference. Furthermore, whereas they
decided a priori how much weight to give high- vs. low-quality studies,
our model estimates these weights empirically based on the noisiness
observed in the different types of data sources.

\section{The Data}
\label{secdata2}
For 199 countries and territories, from 1980 to 2008, we estimate
trends in mean systolic blood pressure (SBP) for adults 25 years of age
and older. We accessed numerous unpublished studies and reviewed
published studies to collate comprehensive data on SBP. We grouped the
199 analysis countries into 21 subregions using the classifications of
the GBD Study. We grouped the subregions into seven merged regions.
Details are given in \citet{SBP2010}.


The primary challenge of this analysis is the fragmented nature and
varying quality of the data, available only from some countries, in
some years and for some age groups. For roughly one-third of all
countries, no data exist at all. Furthermore, many studies cover only
rural or only urban populations. Although a portion of the data comes
from national surveys with sample weights, most data come from
epidemiologic studies not intended to be nationally representative. In
addition, many data sources suffer from small sample sizes.

\section{Why Bayes?}
Given these patterns of data sparsity and missingness, a hierarchical
model is needed to provide inference for all country--year--age triplets
and to account for missingness when aggregating to the regional and
global levels. The hierarchy provides prior distributions that enable
us to borrow strength over time, countries and age, while enforcing
plausible parameter constraints.

In principle, a non-Bayesian hierarchical mixed model is an
alternative, fit by maximum marginal likelihood after integrating over
all the random effects, but the predictive uncertainty would not have
included the substantial uncertainty from hyperparameter estimation.
Furthermore, with 23 hyperparameters, this would have been a
challenging optimization in practice, especially given the parameter
constraints. In addition, it would have been difficult to interpret the
hierarchical model in a non-Bayesian fashion, with mean blood pressure
for a country as a random effect, given that the fixed countries of the
world are not drawn from some large population of possible countries.

MCMC sampling has the added advantage of providing Bayesian imputations
of risk factor levels at any level of aggregation (over age groups,
times, countries, etc.) as a product that the many stakeholders in this
work can use to do their own analyses that easily incorporate
uncertainty; our analysis includes functionals such as the linear
component of blood pressure time trends and the population-weighted,
age-standardized global mean blood pressure level (see Section \ref{secinf}).


\section{The Model}
\label{secmodel2}
Our basic strategy is to fit a Bayesian hierarchical model that
clusters countries within geographical subregions and regions of the
globe, thereby borrowing strength from countries with data. Our
approach treats countries as exchangeable in the absence of other
information, after accounting for covariates. To the basic model we add
smooth time trends and age effects as well as country- and study-level
covariates. We specify a heteroscedastic, multi-component error
structure to account for the fact that not all studies are nationally
representative. Models for women and men are fit separately.

Throughout, bold characters denote vectors and matrices. For each age
group $h$ from study $i$, the model inputs a sample average and a
sample standard deviation of SBP values ($y_{h,i}$ and $s_{h,i}$) as
well as a sample size ($n_{h,i}$). We let $t_i$ denote the year in
which study $i$ was conducted and we use square brackets to denote
group membership such that $j[i]$ is the country $j$ in which study $i$
was conducted. The likelihood is
%
\begin{eqnarray}\label{eqlik}
&&y_{h,i}|a_{j[i]}, b_{j[i]}, u_{j[i],t_i},
\bolds\beta, \gamma _i, e_i, \tau^2_i
\nonumber\\
&&\quad\sim\mathcal N \biggl(a_{j[i]} + b_{j[i]} t_i +
u_{j[i],t_i} \\
&&\hspace*{15pt}\qquad{}+ \mathbf X^\prime_{i} \bolds{\beta}
+ \gamma_i(z_h) + e_i, \frac{s_{h,i}^2}{n_{h,i}}
+ \tau^2_i \biggr).
\nonumber
\end{eqnarray}
$a_j$ and $b_j$ denote the country-specific intercept and linear time
slope for the $j$th country
($j=1,\ldots,J=199$). These intercepts and slopes are modeled
hierarchically, as discussed in Section \ref{subsechier}. $\mathbf
u_j$, a vector of length $T=29$, models smooth nonlinear change over
discretized time ($t=1980,\ldots,2008$) in country $j$ (Section
\ref{subsecnonlin}). The matrix $\mathbf X$ contains study- and
country-level covariates
(Section \ref{subseccovar}). The $z_h$'s are age-group values and the
$\gamma_i(\cdot)$'s are their smoothly-varying study-specific effects;
we describe the flexible age model in Section \ref{subsecage}. 
Finally we add a random effect, $e_i$, to capture study-level
heterogeneity, allowing us to combine data from disparate sources, as
described in Section \ref{subsecssre}.

The likelihood variance has two terms. $s_{h,i}^2/n_{h,i}$ represents
the known sampling uncertainty of mean SBP for a given age group within
a study. We model additional residual variability across age groups
within a study as $\tau_i^2$ (Section \ref{subsecssre}).

\subsection{Linear Components of the Time Trends}
\label{subsechier}
We model the intercepts and slopes in a hierarchical fashion, with each
country-specific intercept, $a_j$, and slope, $b_j$, composed of
country- ($c$), subregion- ($s$), region- ($r$) and global-level ($g$)
components. Letting $k$ index subregions and $l$ index regions, we have
\begin{eqnarray*}
a_j &=& a^c_j + a^s_{k[j]}
+ a^r_{l[j]} + a^g,
\\
b_j &=& b^c_j + b^s_{k[j]}
+ b^r_{l[j]} + b^g.
\end{eqnarray*}
The constituent random intercepts ($a^c$, $a^s$ and $a^r$) and slopes
($b^c$, $b^s$ and $b^r$) each have a normal prior with mean zero and
variance equal to $\kappa^c_a$, $\kappa^s_a$, $\kappa^r_a$, $\kappa
^c_b$, $\kappa^s_b$ or $\kappa^r_b$, respectively. The variance
parameters determine the degree of intercept ($\kappa_a$) and slope
($\kappa_b$) shrinkage performed at the country- ($\kappa^c$),
subregion- ($\kappa^s$) and region-levels ($\kappa^r$). For the
variance parameters, we use a flat prior on the standard deviation
scale (\citeauthor{Gelman2006}, \citeyear{Gelman2006}). We use flat
priors for $a^g$ and $b^g$ as
well. All flat priors were truncated at 0 and 1000.

\subsection{Nonlinear Change in Time}
\label{subsecnonlin}
We also model smooth nonlinear change over time in country $j$ hierarchically:
$\mathbf u_j = \mathbf u^c_j + \mathbf u^s_{k[j]} + \mathbf u^r_{l[j]}
+ \mathbf u^g$,
with each component of the nonlinear trend modeled using a discrete
second-order Gaussian autoregressive prior (\citeauthor{Rue2005},
\citeyear{Rue2005}). In
particular, we model each of the vectors\vspace*{1pt} $\mathbf u^c_j$ ($j=1,\ldots,
J$), $\mathbf u^s_k$ ($k=1,\ldots, K$), $\mathbf u^r_l$ ($l=1,\ldots,
L$) and $\mathbf u^g$ using a normal prior with mean zero and precision
$\lambda_c\mathbf P$, $\lambda_s\mathbf P$, $\lambda_r\mathbf P$ and
$\lambda_g\mathbf P$, respectively. The fixed matrix $\mathbf P$
penalizes second differences.

In this portion of the model, we enforce two constraints to achieve
identifiability. We give the precision parameters a flat prior on the
standard deviation scale (\citeauthor{Gelman2006}, \citeyear
{Gelman2006}), truncating $\operatorname{log}
\lambda\leq15$, as larger values
correspond to essentially no extra-linear temporal variability. We also
enforce orthogonality between the linear and nonlinear components of
the time trends by constraining the mean and slope of each $\mathbf
u^c$, $\mathbf u^s$, $\mathbf u^r$ and $\mathbf u^g$ to be zero.

\subsection{Covariate Effects}
\label{subseccovar}
We include six time-varying, country-level covariates: national income,
national urbanization and four measures of national food availability (namely,
the first four terms from a principal components analysis summarizing the availability
of many food types, e.g., meats, pulses, spices).
We include interactions of income and urbanization
with time because the associations may have changed over time (e.g., as
treatment for high blood pressure became available). We smoothed the
country-level covariates using a triangularly-weighted moving average
with weights decreasing from the year of data collection to the ninth
year prior.

At the study level, we include two covariates to account for potential
bias from data sources that are not representative of national
populations. We
account for potentially time-varying effects of sources that are not
nationally representative. In addition,
we account for differences between study- and country-level
urbanization using an interaction term.

\subsection{Age Model}
\label{subsecage}
Mean SBP generally varies as a nonlinear function of age (\citeauthor
{singh2012}, \citeyear{singh2012}). We model the age effect using
cubic splines with fixed
knots at 45 and 60 years:
\begin{eqnarray*}
\gamma_i(z_h) &=& \gamma_{1i} z_h
+ \gamma_{2i} z_h^2 + \gamma_{3i}
z_h^3 \\
&&{}+ \gamma_{4i} (z_h-45)^3_+
+ \gamma_{5i} (z_h-60)^3_+.
\end{eqnarray*}
We centered the age variable ($z_h$) at 50 years of age to reduce
dependence among model parameters. The $\gamma$'s are modeled as
$\gamma
_{si} = \psi_s + \phi_s\mu_i + c_{sj[i]}$ for $s = 1, \ldots,5$, where
$\mu_i = a_{j[i]} + b_{j[i]} t + \mathbf X_i^\prime\bolds\beta+
u_{j[i],t_i} + e_i$ is the blood pressure level for the 50-year-old age
group. We model the spline coefficients for study $i$ as a linear
effect of the level for this baseline group because blood pressure
tends to increase more sharply as a function of age in countries with
higher SBP levels (\citeauthor{singh2012}, \citeyear{singh2012}).
To this, we add a
country-specific random effect to account for additional
country-specific variation in the age effect, with $c_{sj}|\sigma^2_s
\sim\mathcal N(0, \sigma^2_s)$ and flat priors for the $\sigma_s$'s
(\citeauthor{Gelman2006}, \citeyear{Gelman2006}).

The age model above is continuous in age. However, the blood pressure
means are reported for discrete age groups (e.g., mean SBP for 35--44-year-olds).
As a simplification, we used the midpoint of each age range (e.g., 40
years) as the age value for each data point.

\subsection{Study-Specific Random Effects and Residual Age-by-Study
Variability}
\label{subsecssre}
We account for study-level effects (above and beyond sampling
variability) that are consistent across age groups by including a
study-specific random effect,~$e_i$. We model these random effects as
being normally distributed with a variance that depends on how
representative the study is of the country's population:
\[
\operatorname{Var}(e_i) = \cases{ \nu_w, & if study $i$ is
nationally\cr
& representative with sample weights,
\cr
\nu_u, & if study
$i$ is nationally\cr
& representative without sample\cr
& weights,
\cr
\nu_s, &
if study $i$ is ``sub-national'' (i.e.,\cr
&  covers multiple
provinces/states),
\cr
\nu_c, & if study $i$ is from\cr
& an individual
community.}\hspace*{-4pt}
\]
Exploratory analysis and subject-matter knowledge suggest that even
weighted national studies may have more variability than can be
accounted for by sampling variability because of issues with study
design and quality; this is accounted for through the $\nu_w$ variance
term. We then assume that studies that are increasingly less
representative have increasing random effects variances, imposing the
set of constraints $\nu_w < \nu_u < \nu_s < \nu_c$. The assumption that
we should smooth over (rather than fitting) aberrant data points is
substantiated by the larger-than-expected variability among studies
from country-years in which we have multiple nationally representative
studies with sample weights. 


We also include a variance term for within-study errors (above and
beyond sampling variability) that differ between age groups. As with
the study-specific random effects, we use variance parameters that
differ depending on the representativeness of the study, where $\tau
^2_w, \tau^2_u, \tau^2_s$ and $ \tau^2_c$ are defined in an analogous
fashion to $\nu_w,\nu_w,\nu_s$ and $\nu_s$ and with an analogous
ordering constraint.

\begin{table*}[b]
\tablewidth=335pt
\caption{Decomposition of variability in predictions (\%), with 95\%
credible intervals subscripted}\label{tabvarDecomp}
\begin{tabular*}{\tablewidth}{@{\extracolsep{\fill}}lrrrlr@{}}
\hline
& \multicolumn{1}{c}{\textbf{Country}} & \multicolumn{1}{c}{\textbf{Subregion}}
& \multicolumn{1}{c}{\textbf{Region}} & \multicolumn{1}{c}{\textbf{Globe}}
& \multicolumn{1}{c@{}}{\textbf{Total}}
\\
\hline
& \multicolumn{5}{c@{}}{Female}
\\
Mean & ${}_{26.5}38.6_{51.8}$ & ${}_{3.4}7.6_{14.2}$
& ${}_{16.9}26.9_{38.5}$ & &  ${}_{60.3}73.1_{84.4}$
\\
Lin. trend &  ${}_{2.0}6.8_{15.0}$ & ${}_{0.7}2.7_{6.9}$\hspace*{3.8pt}
& ${}_{3.6}8.5_{15.2}$ & ${}_{0.0}2.6_{10.1}$ & ${}_{11.5}20.6_{31.5}$
\\
Nonlin. trend & ${}_{0.8}4.0_{9.3}$\hspace*{3.8pt} & ${}_{0.1}1.0_{2.8}$\hspace*{3.8pt}
& ${}_{0.1}1.0_{3.2}$\hspace*{3.8pt} & ${}_{0.0}0.3_{1.4}$ & ${}_{1.5}6.3_{13.8}$
\\[6pt]
Total & ${}_{37.4}49.4_{61.0}$ & ${}_{6.2}11.3_{19.8}$
& ${}_{25.2}36.4_{48.5}$ & ${}_{0.1}2.9_{10.7}$ &
\\[4pt]
& \multicolumn{5}{c@{}}{Male}
\\[4pt]
Mean & ${}_{26.4}40.1_{54.1}$ & ${}_{5.7}10.3_{17.1}$
& ${}_{15.2}26.0_{37.3}$ & & ${}_{63.0}76.5_{87.3}$
\\
Lin. trend & ${}_{0.9}3.5_{8.7}$\hspace*{3.8pt} & ${}_{0.3}2.3_{6.9}$\hspace*{3.8pt}
& ${}_{1.1}4.0_{8.2}$\hspace*{3.8pt} & ${}_{0.0}1.5_{7.2}$ & ${}_{5.1}11.3_{19.9}$
\\
Nonlin. trend & ${}_{1.9}6.6_{13.7}$ & ${}_{0.4}1.8_{4.4}$\hspace*{3.8pt}
& ${}_{0.4}2.0_{4.9}$\hspace*{3.8pt} & ${}_{0.1}2.0_{6.4}$ & ${}_{4.3}12.3_{23.2}$
\\[6pt]
Total & ${}_{37.3}50.2_{63.5}$ & ${}_{8.4}14.4_{22.9}$
& ${}_{20.7}32.0_{43.9}$ & ${}_{0.3}3.5_{11.0}$ &
\\
\hline
\end{tabular*}
\end{table*}

\section{Computation}
\label{seccomputation}
We fit the model via Markov chain Monte Carlo (MCMC), using a
combination of conjugate sampling steps and Metropolis--Hastings
updates, with details provided in \citet{SBP2010}. We note that in
hierarchical models there can be strong dependence between parameters
across levels of the model, in particular, dependence of random effects
and their associated variance components. To address this, we jointly
sampled random effects with their hyperparameters (\citeauthor
{Rue2005}, \citeyear{Rue2005},
Section 4.1.2), which greatly improved convergence and mixing.
Finally, we note that while it is possible to analytically integrate
out those parameters in the mean of the normal likelihood whose priors
were also normal, we avoided doing so because it would result in
off-diagonal structure in the covariance of the likelihood, requiring
large matrix manipulations in order to calculate the marginal likelihood.

\section{Model Checking and Inference}
\label{secinf}
We used posterior predictive checks to ensure that we had not omitted
important interactions and used cross-validation to ensure that we had
not overfit our data. In addition, we assessed the sensitivity of our
inference to the inclusion of country-level covariates. All model
checks were reassuring and full results are given in \citet{SBP2010}.
In particular, in the cross-validation our model predicted the
known-but-masked data very well: the 95\% prediction intervals covered
94\% of excluded study mean values for both men and women, consistent
with the expected 95\%.

We draw from the posterior predictive distribution for the mean SBP in
each country, age and year with covariates corresponding to a weighted
national study that represents both urban and rural populations. We
then estimate year- and age-group-specific mean SBP at the subregion
level using a population-weighted average of the mean SBP values for
the countries within the subregion, with analogous estimates for the
regions and globe. We also estimate mean SBP marginalizing over age by
calculating age-standardized values, with weights for each age group
from the World Health Organization standard population. Epidemiologists
are interested in the linear component of the SBP time trends to assess
whether health status has generally been improving. To linearize, at
each iteration we fit a simple linear regression of the country's mean
SBP values against year, collecting the resulting slopes across MCMC iterations.

\begin{figure*}

\includegraphics{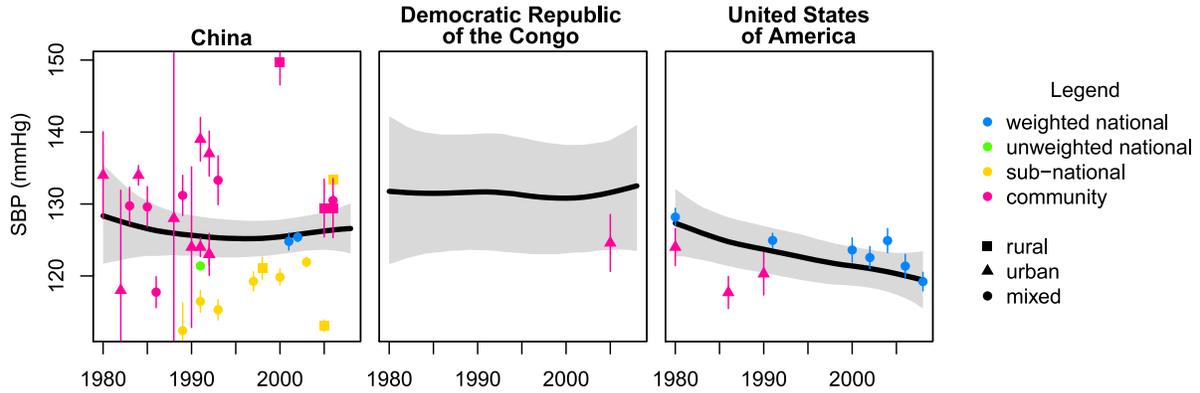}

\caption{Raw data with model fits for 50-year-old females. The solid
line represents the posterior mean, the shaded area the pointwise 95\%
credible interval. The vertical error bars show the 95\% intervals due
to sampling variability ($\pm2s/\sqrt n$).}
\label{figCountryFits}
\end{figure*}

\section{Results}
\label{secresults2}
We additively decomposed the variability in the country--year
predictions for 50-year-olds to understand the variation attributable
to mean and time trends at each of the levels of the hierarchy. For
each country and MCMC iteration, we decomposed the predicted time
series into mean, linear trend and nonlinear trend (residual). We then
decomposed each of these terms into country-specific variation,
subregional variation, regional variation and global variation,
treating country--time points as the units---that is, the subregional,
regional and global terms were averages of the countries within each
subregion, region and globe. This weighting gives greater emphasis to
subregions with more countries than would treating subregions as units
within regions and regions as units within the globe. As can be seen in
Table \ref{tabvarDecomp}, country and region variation predominate,
and cross-country variation is more important than temporal variation.

For females, we note that $v_c$ (the variance of random effects
specific to community studies) is large
(33.0, 27.9--38.8), suggesting
that studies of individual communities do not reflect the country's
mean SBP level accurately. Although $v_w$ (the analogous variance for
nationally representative studies with sample weights) is smaller
(10.8, 6.5--16.0), its magnitude is nonnegligible.
Consistent with this, if we include study-specific variation for
weighted national studies in the variance decomposition above, this
accounts for 22.8\% (13.6--34.4\%) of the variation for females. This
indicates that even weighted national studies, the highest quality
studies in this analysis, may have imperfect study design and quality,
reflected in the anomalous 2004 study in the U.S. (Figure \ref
{figCountryFits}). Similar conclusions hold for males.\looseness=1

\begin{figure*}

\includegraphics{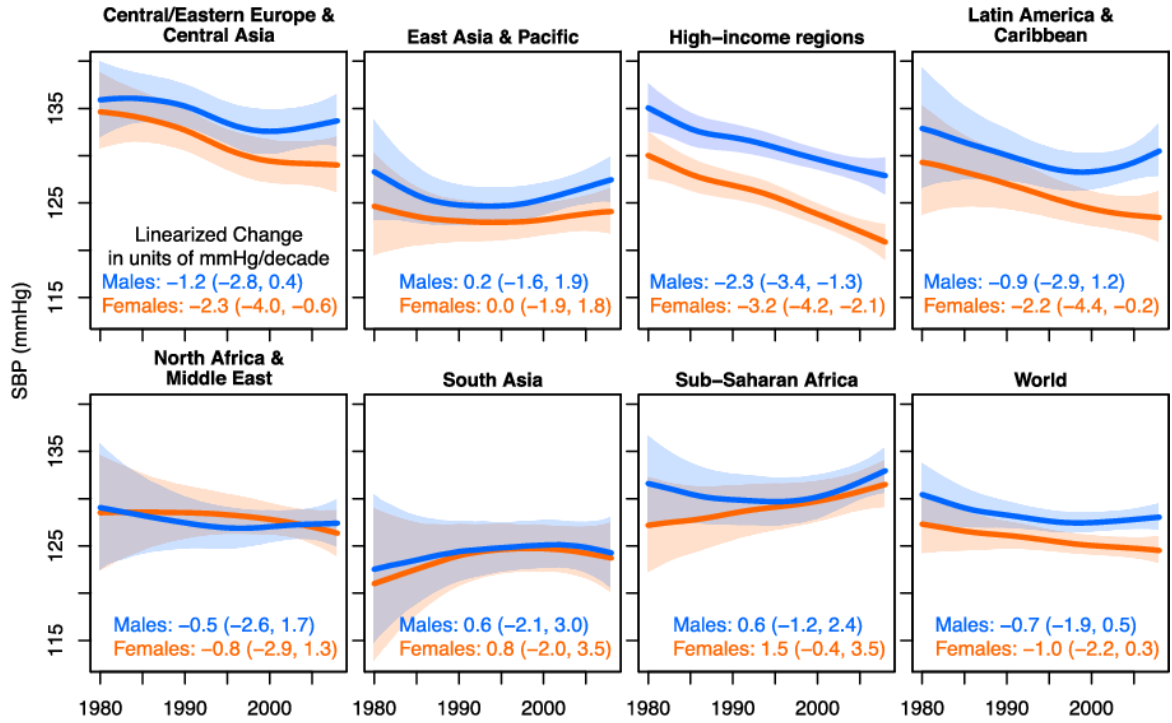}

\caption{Male (blue) and female (orange) trends (estimated separately)
by region. The solid line represents the posterior mean and the shaded
area the pointwise 95\% credible interval. Numerical values are the
estimated linearized time trends.}
\label{figSregionTrends}
\end{figure*}

Figure \ref{figCountryFits} shows example model fits for 50-year-old
females from three countries with differing data density and study
representativeness.

Comparing across subregions in 2008, female SBP was highest in some
east and west African countries, with means of 135 mmHg or greater.
Male SBP was highest in Baltic and east and west African countries,
where mean SBP reached 138 mmHg or more. Men and women in western
Europe had the highest SBP among high-income regions.

Figure \ref{figSregionTrends} shows age-standardized regional and
global trends, highlighting a global transition in which cardiovascular
disease risk factor levels have increased in lower-income regions to
become comparable to---and in places even surpass---those in
high-income regions, in which levels have decreased. A costly epidemic
of high blood pressure in low-income countries may be the most salient
feature of the global cardiovascular risk transition in the coming decades.


\section{Discussion}
\label{secdiscussion2} The results of our analyses using this modeling
strategy were published in a series of four risk-factor-specific papers
in 2011 in \textit{The Lancet} that received press coverage at the
national and global level (including the \textit{Washington Post},
\textit{International Herald Tribune}, \textit{Guardian}, \textit{Times
of India} and \textit{BBC}). The results were used in the WHO Global
Status Report on noncommunicable diseases (NCDs; \citeauthor{WHO2011},
\citeyear{WHO2011}) and The World Health Statistics, and were presented
at the First Global Ministerial Conference on Healthy Lifestyles and
NCD Control. They were used to select ambitious but achievable targets
for cardiovascular disease risk factors for the UN high-level meeting
on NCDs, a task that requires a thorough understanding of past trends.
In addition, our results were used by the US National Academy of
Sciences Panel on \textit{International Health Differences in High
Income Countries} (\citeauthor{Woolf2013}, \citeyear{Woolf2013}) to
understand the role of risk factors for cardiovascular disease in
cross-population health differentials.
Our results were also used to calculate the global burden of disease
attributable to CVD risk factors (\citeauthor{lim2013}, \citeyear
{lim2013}), a calculation which
requires comparable estimates by age, sex, year and country.
Researchers working on non-CVD conditions have also used our results on
CVD risk factors, for example, to examine the role of obesity on
cancers and of maternal obesity on stillbirths in different countries
(\citeauthor{flenady2011}, \citeyear{flenady2011}). Finally, our
close collaboration with leading
global health researchers is helping to place Bayesian methods that
rigorously synthesize fragmentary data at the heart of the conversation
about methods for measures of health status.

While our confidence in the model is bolstered by the cross-validation
results that indicate that our inference reflects the important sources
of variability, there are a number of potential model improvements.
These include further consideration of additional covariates, nonlinear
covariate effects and covariate interactions, including covariate
effects that vary by region. In addition, we would like to have
considered more flexible models for the effects of nonnationally
representative studies and studies representing only rural or urban
populations. While data sparsity led us to assume that a number of
model parameters were constant across region, it would be worthwhile to
investigate allowing the country-level variance components, including
the autoregressive smoothing parameters, to vary by region. Finally,
our model assessment indicated room for improvement in the fitted age
effect in some countries; in particular, age effects may vary with time
beyond our modeled interaction with the overall time-varying level of
mean SBP.

Beyond such model selection issues, we close by noting two important
open issues. First, cross-vali\-dation can only assess our quantification
of predictive uncertainty in relation to the observed data; the
presence of additional variability (beyond sampling variability)
related to shortcomings in study quality in the weighted nationally
representative studies makes it difficult to assess our quantification
of uncertainty in the true country-level trends. Second, we assume that
the presence/absence of data is noninformative; if the studies or
countries represented in the data set are not missing at random, our
results would be biased, with trend estimates affected by data
collection patterns. For example, if countries with more airports tend
to attract both researchers and fast food franchises, then we could be
at risk for overestimating SBP levels.

In summary, efforts to improve global health will depend on reliable
estimates of health status, and many of these estimates will be based
on fragmentary data from disparate sources. The Bayesian paradigm
provides a framework for rigorously combining these data sources to
obtain coherent country-, region- and global-level inference.



%

\end{document}